\documentclass{article}
\usepackage{spconf,amsmath,graphicx,hyperref}
\usepackage{amssymb}
\usepackage{multirow}
\usepackage{booktabs}
\usepackage{color}

\title{Purification Before Fusion: Toward Mask-Free Speech Enhancement for Robust Audio-Visual Speech Recognition}
\name{
\begin{tabular}{c}
Linzhi Wu$^{1}$, Xingyu Zhang$^{2}$\sthanks{Corresponding author. This work was supported in part by the grants from the National Natural Science Foundation of China under Grant 62332019, the National Key Research and Development Program of China (2023YFF1203900, 2023YFF1203903), and sponsored by Beijing Nova Program (20240484513).}, Hao Yuan$^{3}$, Yakun Zhang$^{2}$, Changyan Zheng$^{4}$ \\ Liang Xie$^{2}$, Tiejun Liu$^{1}$, Erwei Yin$^{2}$
\end{tabular}
}
\address{$^{1}$University of Electronic Science and Technology of China, Chengdu, China \\
$^{2}$Defense Innovation Institute, Academy of Military Sciences, Beijing, China \\
$^{3}$Peking University, Beijing, China  \\
$^{4}$High-tech Institute, Weifang, China
}

\begin{document}
\ninept
\maketitle
\begin{abstract}

Audio-visual speech recognition (AVSR) typically improves recognition accuracy in noisy environments by integrating noise-immune visual cues with audio signals. Nevertheless, high-noise audio inputs are prone to introducing adverse interference into the feature fusion process. To mitigate this, recent AVSR methods often adopt mask-based strategies to filter audio noise during feature interaction and fusion, yet such methods risk discarding semantically relevant information alongside noise.
In this work, we propose an end-to-end noise-robust AVSR framework coupled with speech enhancement, eliminating the need for explicit noise mask generation. The framework leverages a Conformer-based bottleneck fusion module to implicitly refine noisy audio features with video assistance. By reducing modality redundancy and enhancing inter-modal interactions, our method aims to preserve speech semantic integrity to achieve robust recognition performance. Experimental evaluations on the public LRS3 benchmark suggest that our method outperforms prior advanced mask-based baselines under noisy conditions.
\end{abstract}

\begin{keywords}
audio-visual speech recognition, speech feature enhancement, noise-robust, multimodal bottleneck Conformer
\end{keywords}

\section{Introduction}
\label{sec:intro}
Audio-visual speech recognition (AVSR) has emerged as a promising solution to overcome the limitations of automatic speech recognition (ASR) systems, especially in noisy conditions. By incorporating visual cues such as lip movements, AVSR leverages complementary multimodal information to improve recognition robustness, attracting considerable attention recently \cite{AfourasCSVZ18,0001PP21a,BurchiT23}. Unlike audio-only ASR, which struggles with degraded performance under challenging acoustic conditions, AVSR exploits visual modality to provide contextual cues that are invariant to acoustic noise, thereby enabling reliable speech transcription even in scenarios with background noise or overlapping speech \cite{HongKYR22,ShiHLM22}. 

Most contemporary AVSR models employ cross-modal attention mechanisms to facilitate effective audio-visual interaction \cite{SterpuSH18,WeiZHD20,SterpuSH20,LiLWQ23,WangGZX24}. Compared to early strategies such as direct feature concatenation, 
cross-attention achieves dynamic alignment and richer multimodal integration.
But, when audio inputs are severely corrupted by noise, the interaction process is prone to disruption.
Noisy audio features can introduce irrelevant or misleading information, forcing the model to 
undertake the entangled tasks of both implicitly \textit{denoising} the input and \textit{extracting} critical speech-related information simultaneously. 
This dual burden might overwhelm the cross-modal interaction module, leading to suboptimal feature fusion.
To alleviate the negative effects of noise, recent studies have proposed custom-designed masking networks to generate noise-reduction masks, which explicitly suppress irrelevant noise components in audio features \cite{HongKYR22,HWANG23,HongKCR23}. These methods aim to suppress noise by selectively focusing on relevant audio information during fusion. Nevertheless, they are typically driven solely by the final AVSR objective, which cannot guarantee the preservation of semantic integrity during this lossy noise-suppression process. 

By contrast, this work presents a purify-then-fuse paradigm without explicit mask generation to achieve noise-robust AVSR. 
The key insight is that prioritizing feature purification before fusion ensures that audio representations fed into cross-modal interaction are semantically complete and noise-free.
To this end, we incorporate an auxiliary audio-visual speech enhancement module based on bottleneck attention into our AVSR framework, which refines audio representations prior to deep multimodal fusion. 
Motivated by \cite{NagraniYAJSS21}, we design an audio-visual bottleneck Conformer to facilitate efficient cross-modal interaction and implicit noise purification.
Guided by audio spectrogram reconstruction objectives, our model tries to extract robust and semantically rich audio representations with the aid of visual cues, mitigating noise interference while retaining essential speech information. 
To our knowledge, this is the first attempt to exploit a multimodal bottleneck Conformer for both efficient cross-modal interaction and reconstruction-based constraints, thus enhancing model resilience to noise.
We evaluate the effectiveness of the proposed method through a series of experiments on LRS3 \cite{lrs3ted18}, the large-scale audio-visual dataset obtained in the wild. The results demonstrate that our model achieves superior performance over previous competitive methods, confirming its robustness under challenging acoustic conditions.

\section{Proposed Method}
\label{sec:method}
\begin{figure}[htb]
  \centering
  \includegraphics[width=0.48\textwidth]{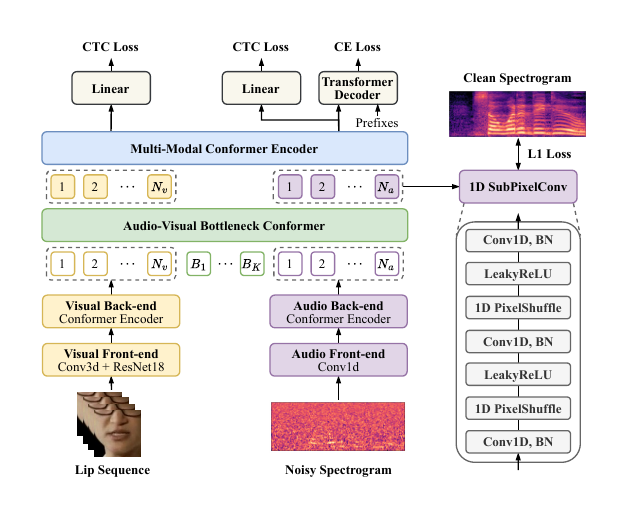}
\caption{Overall architecture of the proposed model.}
\label{fig:framework}
\end{figure}

Fig. \ref{fig:framework} illustrates the proposed noise-robust AVSR framework, which takes video and noisy audio as inputs to perform speech recognition without explicit mask generation. 
This framework integrates a primary AVSR model with a dedicated speech enhancement (denoising) module, which generates refined audio representations from noisy features. Positioned between audio feature extraction and cross-modal fusion, the enhancement module is jointly optimized with the AVSR model. 
These refined audio features are subsequently fed into a cross-modal encoder, where they are fused with visual features to facilitate deep and reliable speech understanding.

\subsection{Feature Extraction}
The video input, denoted as $\mathbf{x}_v$, consists of mouth region-of-interest (RoI) sequences. These are first processed by a 3D convolutional layer with a kernel size of 5$\times$7$\times$7 followed by a 2D ResNet18 \cite{HeResNet16} to extract spatio-temporal features. The resulting features are then fed into a three-layer Conformer encoder \cite{GulatiQCPZYHWZW20}, which captures both local and global visual temporal dynamics, yielding visual features $\mathbf{h}_v \in \mathbb{R}^{N_v \times d}$. The audio input, denoted as log mel-spectrogram $\mathbf{x}_a$, is first processed by two subsampling 1D convolutional layers to reduce the temporal dimension to match the frame rate of the visual features. A similar Conformer is then applied to encode temporal features, producing audio features $\mathbf{h}_a \in \mathbb{R}^{N_a \times d}$. Here $N_v$ and $N_a$ are total frames of video and audio respectively, and $d$ is the feature dimension. Each modality frame refers to a token in our work.

\subsection{Audio-Visual Bottleneck Conformer}
To enhance inter-modal interaction and compress redundant modal information, we introduce an audio-visual bottleneck Conformer (AVBC) with $L$ layers, inspired by \cite{NagraniYAJSS21}. 
The AVBC employs a small set of learnable bottleneck tokens $\mathbf{b}^{0}=[b_1, b_2, \dots, b_K]$, $b_i\in\mathbb{R}^d$ ($K\ll N_a, N_v$). For each modality, cross-attention is computed between its feature sequence and the bottleneck tokens. Formally for layer $l$, we calculate token representations as follows:
\begin{equation}
    \begin{split}
        \mathbf{h}_v^{l+1} \| \mathbf{b}_{v}^{l+1} &= \mathrm{Conformer}(\mathbf{h}_v^{l} \| \mathbf{b}^{l}; \theta_v), \\
        \mathbf{h}_a^{l+1} \| \mathbf{b}_{a}^{l+1} &= \mathrm{Conformer}(\mathbf{h}_a^{l} \| \mathbf{b}^{l}; \theta_a), \\
        \mathbf{b}^{l+1} &= (\mathbf{b}_{v}^{l+1} + \mathbf{b}_{a}^{l+1}) / 2,
    \end{split}
\end{equation}
where $\theta_v$ and $\theta_a$ are the parameters of the visual Conformer and the audio Conformer, respectively. ``$\|$" denotes the concatenation operation of the latent tokens. Note that the convolution module within the Conformer block is only used for modality tokens excluding the bottleneck tokens.
The iterative bottleneck attention computation across $\mathbf{h}_a$ and $\mathbf{h}_v$ ultimately yields refined representations $\mathbf{z}_v = \mathbf{h}_v^{L}$ and $\mathbf{z}_a = \mathbf{h}_a^{L}$.
Since all cross-modal attention flow must pass through these bottleneck tokens, such tight ``fusion" bottlenecks force the model to condense modality-specific information and share only essential content.
This design allows the visual modality to guide the purification of noisy audio features in a computationally efficient manner, which reduces cross-modal attention computations from $\mathcal{O}((N_a + N_v)^2)$ to $\mathcal{O}((K + N_a)^2) + \mathcal{O}((K + N_v)^2)$. 



\subsection{Speech Feature Enhancement}
To refine noisy audio features into fusion-ready representations, we reconstruct the clean mel-spectrogram $\hat{\mathbf{x}}_a$ from $\mathbf{z}_a$ via a dedicated 1D sub-pixel convolution layer \cite{ShiCHTABRW16}, which efficiently upscales audio feature maps to produce the final spectrogram output \footnote{For the spectrogram decoder, our preliminary experiments have shown that the designed lightweight sub-pixel convolution module delivers performance comparable to the widely-used transposed convolution network, while requiring fewer parameters and offering higher computational efficiency.}.
The reconstruction loss is defined as the L1 distance between the mel-spectrogram of the clean audio and the reconstructed mel-spectrogram:
\begin{equation}
    \begin{split}
        \mathcal{L}_{\text{recon}} = \| \hat{\mathbf{x}}_a - \mathbf{x}_a^{\text{clean}} \|_{1},
    \end{split}
    \label{eq:recon}
\end{equation}
Inspired by the perceptual loss used for image processing \cite{JohnsonAF16}, we further introduce a spectrogram-based perceptual loss. Unlike reconstruction loss, which only focuses on raw spectrogram similarity, this loss function captures perceptual discrepancies by minimizing the L2 distance between high-level feature maps of the target and reconstructed mel-spectrograms, where the feature maps are typically obtained by an optimized feature extractor:
\begin{equation}
    \begin{split}
        \mathcal{L}_{\text{percep}} = \| \mathcal{F}(\hat{\mathbf{x}}_a) - \mathcal{F}(\mathbf{x}_a^{\text{clean}}) \|_2^2,
    \end{split}
    \label{eq:cyc}
\end{equation}
where $\mathcal{F}$ is our custom audio front-end in our work, achieving a cycle-consistency regularization. 
The whole speech feature enhancement process is guided by these two loss functions:
\begin{equation}
    \begin{split}
        \mathcal{L}_{\text{enhance}} = \alpha_1\mathcal{L}_{\text{recon}} + \alpha_2\mathcal{L}_{\text{percep}},
    \end{split}
\end{equation}
with the first loss term providing stability (avoid drift in scale) and the second one driving speech intelligibility. $\alpha_1$ and $\alpha_2$ denote the weighting factors for each loss term.


\subsection{Fusion and Speech Recognition}
The cross-modal fusion encoder is not required to dedicate substantial effort to denoising; instead, its primary role focuses on integrating the purified audio with visual information.
The refined representations $\mathbf{z}_v$ and $\mathbf{z}_a$ are concatenated along the temporal dimension and processed by a Conformer encoder 
to fully capture intra-modal and inter-modal temporal dependencies \cite{LiLWQ23},
yielding fused features $\mathbf{f}_v$ and $\mathbf{f}_a$ as follows: 
\begin{equation}
    \begin{split}
        \mathbf{f}_a \| \mathbf{f}_v = \mathrm{Conformer}(\mathbf{z}_a \| \mathbf{z}_v; \theta_f),
    \end{split}
\end{equation}
where $\theta_f$ is the encoder parameter.
These are fed into the CTC \cite{GravesFGS06} projection layer and Transformer decoder \cite{VaswaniSPUJGKP17} for speech recognition. Given the target text sequence $\mathbf{y}$, the recognition objective function is defined as a hybrid CTC/attention loss \cite{hori-etal-2017-joint,WatanabeHKHH17}:
\begin{equation}
    \begin{split}
        \mathcal{L}_{\text{AVSR}} &= \lambda \mathcal{L}_{\text{ctc}} + (1-\lambda)\mathcal{L}_{\text{att}}, \\
        \mathcal{L}_{\mathrm{ctc}} &= -\log p_{\mathrm{ctc}}(\mathbf{y}|\mathbf{f}_v) -\log p_{\mathrm{ctc}}(\mathbf{y}|\mathbf{f}_a), \\ 
        \mathcal{L}_{\mathrm{att}} &= -\log p_{\mathrm{att}}(\mathbf{y}|\mathbf{f}_a),
    \end{split}
\end{equation}
where $\lambda$ balances the relative weight between the CTC and cross-entropy attention-based losses. The CTC projection layer is shared across modalities. 

The overall optimization objective combines recognition and enhancement losses:  $\mathcal{L}_{\text{total}} = \mathcal{L}_{\text{AVSR}} + \mathcal{L}_{\text{enhance}}$.
This speech enhancement module is jointly trained with the AVSR modules. Accordingly, the loss function backpropagates through the enhancement module, compelling it to produce audio representations that are optimal for speech transcription, not merely for spectral fidelity.

\section{Experiments}

\subsection{Data Processing}
\noindent\textbf{Dataset.}
We evaluate the proposed method on the large-scale audio-visual benchmark LRS3 \cite{lrs3ted18}, which consists of about 439 hours of TED and TEDx English talks sourced from YouTube \footnote{\url{https://www.robots.ox.ac.uk/\~{}vgg/data/lip_reading/lrs3.html}}.
The training set is divided into two partitions: \textit{pretrain} (408 hours) and \textit{trainval} (30 hours), both of which are transcribed at the sentence level. The \textit{pretrain} part differs from \textit{trainval} in that the duration of its video clips span a much broader range. The standard \textit{test} set (0.9 hours) is used for evaluation.

\noindent\textbf{Processing.}
For the video stream, mouth RoIs are cropped from the aligned talking face video sequence at 25Hz, resized to 96$\times$96, converted to grayscale, and normalized to [0, 1]. During training, we apply random cropping with crop size 88$\times$88, horizontal flipping (probability 0.5) and adaptive time masking \cite{MaPP22} to augment the video data. We apply center crop at inference time. For the audio stream, raw waveforms are resampled at 16kHz and transformed into 80-dimensional log mel-spectrograms over windows of 25 ms strided by 10 ms. To synthesize noisy audio inputs for training, we select white, pink, factory1, factory2 and babble noises from the NOISEX-92 database \cite{VargaS93}, and extract additional overlapping \textit{speech} noise samples from the training data. These noise samples are then additively mixed with clean waveforms at a signal-to-noise ratio (SNR) level in \{-7.5, -2.5, 2.5, 7.5, 12.5, 17.5\} dB. For each training sample, we uniformly randomly select either a clean or noisy speech.

\subsection{Experimental Settings}
\noindent\textbf{Model Setup.}
Each Conformer encoder within our framework for each modality consists of 3 blocks with 4 attention heads, hidden dimensions of 512, feed-forward dimensions of 2048, and a convolution kernel size of 31. We leverage a standard Transformer decoder for prediction and the decoder comprises 6 layers, 4 attention heads, hidden dimensions of 512, and feed-forward dimensions of 2048. 
Following \cite{0001PP21a,HongKCR23}, the visual front-end is initialized using a pre-trained model on LRW \cite{ChungZ16}.
The learnable bottleneck tokens are initialized using a Gaussian with a mean of 0 and a standard deviation of 0.02, and the number of tokens is set to 4.

\noindent\textbf{Training Setup.}
We use the AdamW optimizer \cite{LoshchilovH19} to update model parameters for 70 epochs with a batch size of 16. 
The initial learning rate is set to 0.001, adjusted by a cosine annealing schedule with a linear warm-up. $\lambda$ is set to 0.1 as suggested in \cite{MaPP22}, and $\alpha_1$ = $\alpha_2$ = 0.1 in our work. 
For the training stability of the entire framework, we adopt a curriculum learning strategy. We first train the model for 20 epochs with the primary AVSR objective using relatively high SNRs (\textit{i.e.}, 7.5-17.5dB), which focuses on establishing robust multimodal alignments and avoids early divergence induced by high-noise inputs. Then, we extend training to the full SNR range and incorporate auxiliary speech enhancement objectives for joint learning, enabling refined feature purification across various SNRs. 

\noindent\textbf{Testing Setup.}
We utilize the model averaged over the last 10 checkpoints for evaluation following \cite{MaPP22}. During inference, decoding is performed with a left-to-right beam search of width 40 on the Transformer decoder, and following \cite{PrajwalAZ22}, we exploit an off-the-shelf pre-trained GPT-2 language model \cite{radford2019language} for beam rescoring. Following prior literature \cite{AfourasCSVZ18,0001PP21a}, 
we report word error rate (WER) on LRS3 under both clean (SNR=$\infty$) and noisy conditions with varying SNR levels.
A lower WER indicates better recognition performance, while a lower SNR corresponds to a higher noise level.

\begin{figure}[t]
    \centering
    \includegraphics[width=1\linewidth]{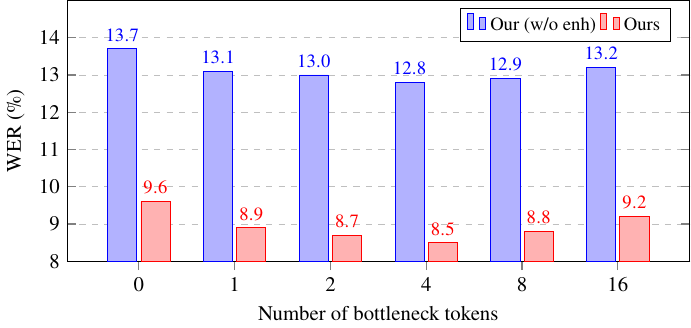}
    \caption{Performance comparison of different token numbers at the first layer of our bottleneck Conformer under -5dB SNR babble noise. ``w/o enh" indicates the model without speech enhancement.}
    \label{fig:bottleneck}
\end{figure}

\begin{table}[t]
\centering
\caption{Effects of the reconstruction loss ($\mathcal{L}_\text{recon}$) and the cycle-consistency loss ($\mathcal{L}_\text{percep}$) on WER, where $\mathcal{L}_\text{percep}$ is applied to the audio front-end and a pre-trained Whisper encoder, respectively.}\label{tab:enhance}
\begin{tabular}{cccc}
\toprule
\textbf{Baseline} & $\mathcal{L}_{\text{recon}}$ &  $\mathcal{L}_{\text{percep}}$ (Audio front-end)  & \textbf{WER (\%)}  \\ \midrule
   $\checkmark$  &  --  &  --  &  12.8  \\
   $\checkmark$  &  $\checkmark$  &  --  &  11.2  \\
    $\checkmark$  & --  &  $\checkmark$  &  10.8  \\
   $\checkmark$  &  $\checkmark$  &  $\checkmark$  &  \textbf{8.5} \\ \midrule\midrule
\textbf{Baseline} & $\mathcal{L}_{\text{recon}}$ &  $\mathcal{L}_{\text{percep}}$ (Whisper)  & \textbf{WER (\%)} \\   \midrule
    $\checkmark$  &  --  &  --  &  12.8 \\
   $\checkmark$  &  $\checkmark$  &  --  &  11.2  \\
    $\checkmark$  & --  &  $\checkmark$  &   9.5  \\
   $\checkmark$  &  $\checkmark$  &  $\checkmark$  &  \textbf{7.9}  \\
\bottomrule
\end{tabular}
\end{table}

\begin{table*}[t]
\centering
\caption{WER (\%) comparisons with other competitive methods within different noise levels. }\label{tab:baselines}
\begin{tabular}{cccccccc}
\toprule
\multirow{2}{*}{\textbf{Method}} & \multicolumn{6}{c}{\textbf{SNR (dB)}}  & \multirow{2}{*}{\textbf{Avg.}} \\\cmidrule{2-7}
& $\infty$ & \textbf{15} & \textbf{10} & \textbf{5} & \textbf{0} & \textbf{-5} &  \\ \midrule 
EG-Seq2Seq \cite{XuLGW20}  &  6.8  &  3.3  &  3.9  &  5.8  &  12.2  &  27.3  &   9.9 \\
Conformer \cite{0001PP21a}  &  3.2  &  3.6  &  5.4  &  8.3 &  14.6 &  22.3  &  9.6      \\
V-CAFE \cite{HongKYR22} &  2.9  &  3.0  &  4.0  &  8.4  &  12.5  &  19.3    &   8.4    \\
Joint AVSE-AVSR \cite{HWANG23}  & 2.0  &  2.4 &  2.9 &  4.1 &  8.0  &  19.4  &  6.5  \\
AV-RelScore \cite{HongKCR23} &  2.8  & 2.9  & 2.9  &  3.3 &  4.8  &  9.0   &   4.3  \\ 
Ours (w/o enh) &  2.3   &  3.1  &  3.8 &  4.6  &  6.7  &   12.8  &   5.6          \\
Ours  &   \textbf{2.1}    &  \textbf{2.4}  &  \textbf{2.6}  &   \textbf{3.2}  &  \textbf{4.5}  &   \textbf{8.5}  &   \textbf{3.9}  \\
\bottomrule
\end{tabular}
\end{table*}

\begin{table}[t]
\centering
\caption{WER (\%) comparisons of different methods under clean audio (SNR=$\infty$) and overlapped speech (SNR=-5dB) inputs, with or without video assistance.}\label{tab:inputs}
\begin{tabular}{ccccc}
\toprule
\multirow{2}{*}{\textbf{Input Condition}} & \multicolumn{2}{c}{\textbf{Clean audio}} & \multicolumn{2}{c}{\textbf{Overlap. speech}} \\  
& w/ vid      & w/o vid     & w/ vid          & w/o vid   \\ \midrule
Unified-Attention \cite{LiLWQ23}  &  2.4    &  2.7  &  11.3  &  27.4  \\
Our (w/o enh)   &   2.3      &  2.5   &   10.7    &   25.9      \\
Ours     &     \textbf{2.1}      &  \textbf{2.2}       &  \textbf{9.6}    &  \textbf{24.6}    \\ 
\midrule
\end{tabular}
\end{table}

\subsection{Results and Analysis}

\subsubsection{Influence of the number of bottleneck tokens}
The bottleneck tokens serve a key role in our AVBC fusion structure. We thus conduct experiments to evaluate how their quantity influences the model's final prediction performance under -5dB babble noise. We keep the number of bottleneck tokens to be much smaller than the total latent units per modality, explicitly ensuring only necessary information resides in these bottleneck tokens. Specifically, we set the number of tokens to 0, 1, 2, 4, 8, and 16 respectively for comparison, where ``0'' means that the two modalities skip the bottleneck and directly engage in cross-attention. As depicted in Fig. \ref{fig:bottleneck}, we can observe that the model delivers relatively favorable performance when the number of tokens is set to 4. The results imply that too few tokens may hinder sufficient exchange of essential inter-modal information, while too many could compromise the model’s ability to prioritize only essential content transmission.

\subsubsection{Effectiveness of the speech enhancement}
The audio spectrogram reconstruction within our framework is optimized with two loss functions, \textit{i.e.} the spectrum-based term ($\mathcal{L}_{\text{recon}}$ in Eq. \ref{eq:recon}) and the perception-based term ($\mathcal{L}_{\text{percep}}$ in Eq. \ref{eq:cyc}). We try to assess the effect of each individual loss on final performance. Besides the audio front-end, we also exploit a pre-trained ASR encoder, Whisper \cite{RadfordKXBMS23} \footnote{\url{https://github.com/openai/whisper}} medium with fixed parameters, to obtain $\mathcal{L}_{\text{percep}}$. Table \ref{tab:enhance} presents the ablation results under -5dB SNR babble noise, and we find that the combination of these two types of losses can consistently lead to better performance. As our primary objective is AVSR performance not just for clean-sounding audio restoration, the $\mathcal{L}_{\text{percep}}$ term tends to contribute more to recognition performance by extracting high-level semantic structures of the enhanced audio to improve intelligibility. Compared with the audio front-end, the Whisper can better encourage the enhanced output to preserve rich phonetic and linguistic information that matters for decoding results. 
But, due to its higher computational overhead, it significantly slows training. Given that our primary goal is robust AVSR -- not audio enhancement in isolation -- we opt for the audio front-end in this work to maintain training efficiency while still achieving effective gains.

\subsubsection{Noise robustness evaluation}
To examine the noise robustness of our model against prior competitive baselines, we perform comparative experiments across acoustic conditions: clean audio (SNR=$\infty$) and noisy settings with an SNR range of -5 to 15 dB. As shown in Table \ref{tab:baselines}, our proposed method outperforms these advanced noise-robust methods that require explicit noise-reduction masks (\textit{e.g.}, AV-RelScore \cite{HongKCR23}, which incorporates a scoring module to suppress unreliable representations) by achieving a lower average WER. Particularly, as SNR decreases, the performance gap between our method and these baselines widens. Moreover, our full model reduces WER by 1.7\% compared to the ablation variant without speech feature enhancement objectives. The results show the effectiveness of our method in suppressing noisy audio representations without relying on explicit noise mask generation.

\subsubsection{Experiments on various input conditions}

Since cross-modal attention operations are decoupled from the modal encoders, our AVSR framework generalizes to single-modality input sequences, aligning with \cite{LiLWQ23,LiLWQ24}. To investigate the model's robustness under diverse input conditions, we further conduct evaluations covering clean audio and overlapped speech, each assessed with and without video input. Overlapped speech is generated by randomly selecting a sample from the test set that is distinct from the input audio and overlapping it with the input audio at an SNR of -5dB. During testing, the two distinct speech signals are played simultaneously, while the input video only provides visual cues for the target speech. Notably, the bottleneck tokens within the AVBC still interact with the audio sequence even when video input is disabled. As shown in Table \ref{tab:inputs}, the presence or absence of video input exerts minimal influence on the final performance under clean audio conditions. However, when overlapped speech is used as the audio input, the visual modality becomes crucial for effectively ``selecting'' the target speech from the mixed speech. Furthermore, our model (without speech enhancement) performs better than the baseline in \cite{LiLWQ23} under overlapped speech scenarios. This advantage likely stems from the bottleneck tokens preserving modality-shared features, which serve as complementary cues for the corrupted audio.


\section{Conclusion}
In this paper, we propose a noise-robust AVSR architecture that integrates AVSR with speech enhancement, achieving noise robustness without resorting to explicit noise mask generation. We take advantage of the bottleneck Conformer and audio reconstruction objectives to effectively refine noisy audio representations with video assistance, while preserving speech semantic integrity to facilitate subsequent cross-modal fusion. 
Experimental results on the LRS3 dataset show superior noise robustness compared to mask-based baseline methods, 
validating the efficacy of implicit noise suppression enabled by bottleneck-driven cross-modal fusion.
\bibliographystyle{IEEEbib}
\bibliography{strings,refs}

\end{document}